*Review*

# Automated detection of Alzheimer disease using MRI images and deep neural networks- A review

**Narotam Singh[1], Patteshwari.D[1], Neha Soni[2] and Amita Kapoor[3,*]**

[1]Department of Cognitive Neurosciences, School of Life Sciences, JSS Academy of Higher Education and Research, Mysuru, Karnataka, India; narotam.singh@gmail.com
[2]University of Delhi South Campus, Delhi, India
[3]SRCASW, University of Delhi, Delhi, India
* Correspondence: dr.amita.kapoor@ieee.org

**Abstract:** Early detection of Alzheimer's disease is crucial for deploying interventions and slowing the disease progression. A lot of machine learning and deep learning algorithms have been explored in the past decade with the aim of building an automated detection for Alzheimer's. Advancements in data augmentation techniques and advanced deep learning architectures have opened up new frontiers in this field, and research is moving at a rapid speed. Hence, the purpose of this survey is to provide an overview of recent research on deep learning models for Alzheimer's disease diagnosis. In addition to categorising the numerous data sources, neural network architectures, and commonly used assessment measures, we also classify implementation and reproducibility. Our objective is to assist interested researchers in keeping up with the newest developments and in reproducing earlier investigations as benchmarks. In addition, we also indicate future research directions for this topic.

---

## 1. Introduction

Alzheimer's disease (AD) is the most common cause of dementia worldwide; it is a degenerative neurological disorder that is progressive; due to it, the brain cells die slowly. Forgetfulness and memory loss of recent events are the early symptoms of Alzheimer's disease. And as the disease progresses, it affects all cognitive functions, eventually making the patient completely dependent for even the basic functions of daily life. It is estimated that worldwide there are as many as 55 million people suffering from Alzheimer [1], and it is expected that the number will increase by 10 million every year. If unmanaged, this results in an expensive public health burden in the years to come.

There is no cure for Alzheimer's; if detected early, you can lead an everyday life with lifestyle changes and manage it for a considerable length of time. The first step in the automated detection of Alzheimer's is feature selection- choosing the right features to feed the deep learning model. In this paper, we have considered papers focusing on (Magnetic resonance Imaging (MRI) data as the input. MRI images are often 3D, and thus result in large feature space, making feature selection an essential component. The traditional methods of extracting features are frequently constrained by some a priori information, which means that they can only extract certain aspects that relate to a particular application. Therefore, many researchers have shifted to using deep learning for feature selection as well. There are two advantages of using deep learning techniques for feature selection. First, it can be implemented to automatically find features for a certain application based on a given data source. Second, it has the ability to find novel features that are suitable for particular applications, features that researchers have never found before but which are suited for those applications.

In this review, we will concentrate on the most recent emergent deep learning, which is represented by various deep learning algorithms [2]. Deep learning has achieved tremendous success in recent years, thanks to the collection of big data from the Web, the parallel processing ability of graphics processing units (GPUs), and the new convolutional neural network family, for a variety of applications such as image classification [3, 4], object detection [5], time series prediction [6, 7], etc.

Due to the numerous new breakthroughs in this field, it is tough for a beginner to keep up with the most recent advancements. To address this issue, we provide a summary of the most recent advancements in deep learning algorithms for automated AD detection, focusing on those that have been written during the past two years. In addition, we give the trend of each phase of the prediction workflow over the past two years, allowing newcomers to avoid wasting time on outmoded technologies.

In addition, we pay particular attention to the execution and reproducibility of earlier investigations, which are sometimes overlooked in comparable studies. The list of open data and code from published articles will not only assist readers in verifying the validity of their conclusions, but also allow them to deploy these models as baselines and conduct fair comparisons on the same datasets. Based on our assessment of the surveyed publications, we attempt to identify future research directions that will assist the readers in determining their next step.

Our major contribution can be summed up as follows:

1. We summarize the latest progress of applying deep learning techniques to AD prediction, especially those which only appear in the past two years (2020-2022).
2. We specifically considered MRI for AD prediction.
3. We give a general workflow for AD prediction, based on which the previous studies can be easily classified and summarized. And future studies can refer to the previous work in each step of the workflow.

This survey's remaining sections are organized as follows: Section 2 presents relevant material; The third section provides an overview of the papers covered; Section 4 outlines the key findings in each step of the prediction workflow; Section 5 discusses implementation and reproducibility; and Section 6 identifies potential future research directions. Section 7 concludes this survey.

## 2. Related Work

Computer aided Automated detection of AD has been of interest for almost a decade now. The reviews prior to 2017, [8], focused only on machine learning methods, with special emphasis to feature selection. However, later work found that end-to-end deep learning has good performance and techniques like transfer learning can boost the performance.

The review paper by Shoukry et al. [9] considered papers published between 2003-2019. According to their mini review- while interest was growing in using the image dataset, the work was not very relevant since most of the datasets considered were of AD patients and hence of little use in early detection of AD.

Tanveer et al. [10] work considered in total 165 papers published between 2005-2019. They focused on three major machine learning techniques- support vector machines (SVM), artificial neural networks (ANN), and deep learning (DL). According to their analysis, SVM based techniques were more robust and they hoped that deep learning techniques would hopefully give better results in the future.

Yang and Mohammed [11] reviewed 32 papers published between 2017-2020. In their work, they considered studies using multi-channel convolutional neural networks (CNNs), pre-trained CNNs, CNN/recurrent neural networks (RNN) autoencoders, and combination of CNN and RNN, beside traditional machine learning approaches like SVM. The datasets considered was mainly Alzheimer's disease neuroimage initiative (ADNI) data. The authors pointed out that the results of the papers cannot be relied upon since none dealt with the issue of class imbalance.

Ebrahimighahnavieh et al. [12], considered papers from 2013 onwards. According to their review, preprocessing of brain scans played a crucial role, multi-modal studies performed better than single-modal ones, data augmentation gives mixed results, a balanced dataset is recommended, and transfer learning outperforms other techniques.

Gao and Lima [14] in their 2021 review paper considered 12 papers in the period 2017-2021. The images included both MRI images and Positron Emission Tomography (PET) scans. The strength of the review paper was that it provided advantages and weaknesses of the approaches covered for automated AD detection. According to them, end-to-end deep learning with CNN is the preferred method as compared to manual extraction of the features, with interest growing in employing the transfer learning approach.

Afzal et al. [15] divided the papers reviewed into three major categories: SVM based techniques (13 papers); machine learning based techniques (which included techniques like Bayesian networks, logistic regression, CNN, and RNN) (16 papers); and transfer learning-based approaches (10 papers). The review considered preprocessing software as well as the deep learning frameworks employed for the work. According to them, transfer learning approaches were most successful due to the limited size of available neuro-image datasets.

Kumar et al [17], considered articles published between 2010-2020. They considered different aspects like data source and format, characteristics, methodologies, and research topics. According to their results, the majority of contemporary research on AD has concentrated on forecasting the development of the illness using standardized, publicly available multimodal datasets that include neuroimaging and some non-imaging clinical data. Neurobehavioral status exam scores, patient demographics, neuroimaging data, and laboratory test values are the non-imaging clinical data utilized most frequently for predictive modeling.

Borchert et al, [16], reviewed 252 papers with a publication time between 2005-2021. The review focused on countries doing the research in the area and the datasets used, with ADNI being the most used dataset in their samples (70.6%). They covered structural MRI, fMRI, PET/SPECT and multiple modalities. Despite the increase in research in the area, the authors had reservations about its clinical integration.

Grueso and Sobera [18], considered 116 papers published between 2010-2021. Their major contribution was analyzing the bias in proposed solutions.

We can see that the review papers spanned almost two decades, from 2003 to 2021. The least number of papers reviewed was 12, and while Borchert covered 252 papers spanning from 2005-2021. Almost all the review papers suggest that deep learning gives better results. Thus, we, in this paper, focused only on deep learning techniques published between January 2020- January 2022, covering a span of 24 months, with a total of 244 papers in the initial search. Following the systematic elimination as described in Figure 1, we were left with 60 papers. That is roughly 2.5 papers per month.

## 3. Overview

In this section, we give an overview of the papers reviewed in this study. All the works are searched and collected from Google Scholar using its SERP API, with searching keywords: "Alzheimer" "fMRI" and "deep learning". We have covered the papers published in the last two years (January 2020-January 2022). Initially, 244 articles appeared in the Google Scholar search for two years. Out of these, 103 articles are selected based on the title and abstract. Out of 103 articles, 46 articles were rejected for one of two reasons:

1. Unavailability of full article
2. Book chapters, thesis, and review papers are rejected

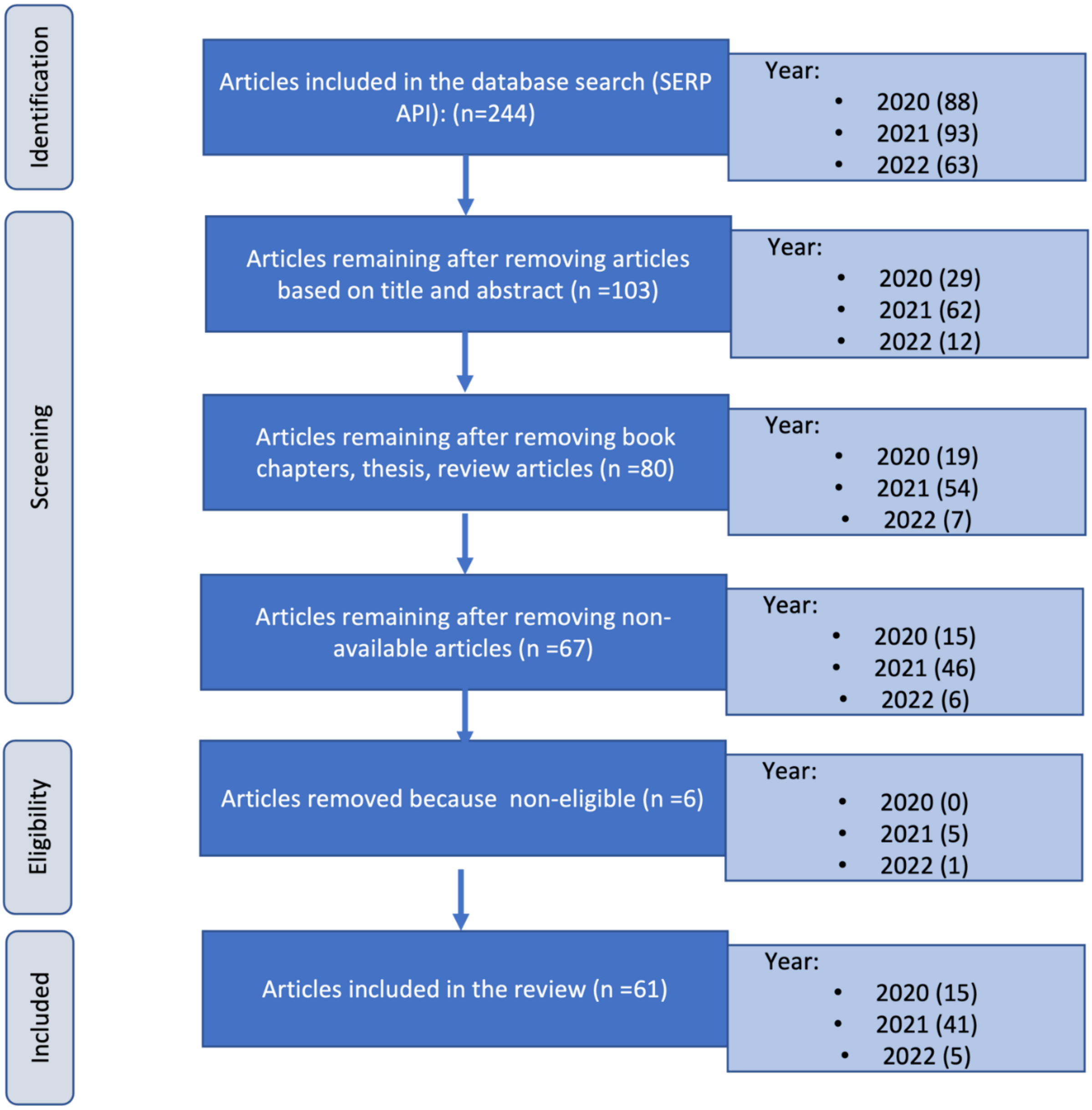

Figure 1 Article selection criteria for the survey

Finally, 61 articles were selected and are included in the survey. Figure 1 shows the detailed steps followed to select 61 papers from 244 papers in search list. In total, we cover 40 journal papers, 19 conference papers, and 1 preprint paper. The preprint article comes from the well-known e-print archive website arXiv.org, and we cover these publications to keep researchers updated with the most recent developments. The top source journals sorted by the number of papers we covered in this study are shown in Figure 2.

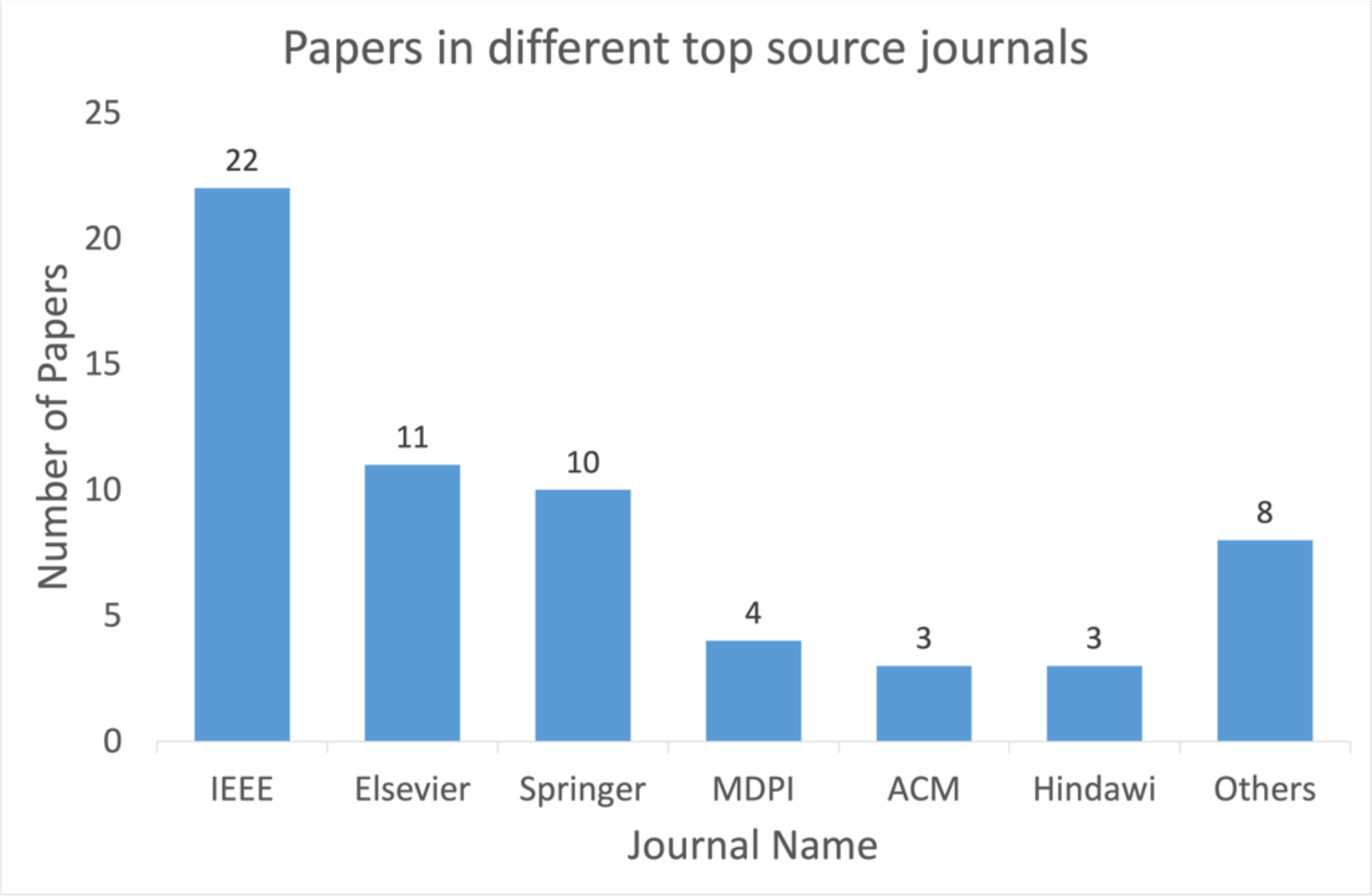

Figure 2 List of top source journals and the number of papers we covered in this survey

## 4. Prediction Workflow

The research papers considered use a varied combination of data sources, and models. Therefore, in this section, we summarize the general workflow with four steps that most of the studies follow: Brain Imaging Techniques, Alzheimer Datasets, Prediction Models and Model Evaluation. In this section, we will discuss each step separately.

### *Brain Imaging Techniques*

For deep learning, data is a crucial component. Thus, here we list different brain imaging techniques used to get brain images.

#### Structural Magnetic Resonance Imaging (sMRI)

MRI is a non-invasive (non-anatomical) imaging (illustration) technique used to evaluate the structural integrity of the affected brain region. Brain scans are performed with MRI scanners. The imaging target is bombarded with magnetic rays during the MRI process. MRI is widely used for the early diagnosis of Alzheimer's disease. Due to the availability of open-source databases like ADNI and Open Access Series of Imaging Studies (OASIS), the number of MRI-based studies has increased over the years. This has also had an effect on the studies that utilize these databases to detect disease progression and improve analytical studies. Using MRI, it is also possible to observe the disease's effect on the subject's spatial domain and temporal region, which creates patterns.

Brain tissue degeneration in Alzheimer's disease patients and the transition from mild cognitive impairment (MCI) to Alzheimer's disease can occur more rapidly than in healthy individuals, according to MRI-based research. In this regard, the concerned domain could be the detection of subjects with MCI, which is possible through the early detection of AD. Multiple studies indicate that MRI is not a foolproof method for detecting AD in its early stages. The degeneration in the hippocampal region is easily identifiable in Alzheimer's disease patients compared to non-AD patients. As in many cases, the brain damage is not solely attributable to AD; therefore, identifying these subtle distinctions can be crucial. Over time, however, MRI-based studies (automatic MRI) demonstrated encouraging improvement. Currently, MRI-based techniques are widely used for the early detection of AD due to the widespread availability of MRI devices.

#### Functional Magnetic Resonance Imaging (fMRI)

Functional MRI is also a noninvasive procedure that aids in the diagnosis of AD-related dysfunction. fMRI also permits the observation of oxygen absorption during resting and active states to establish an activity pattern. Thus, brain activity during various states can be assessed. fMRI extracts data from each brain region to aid in the diagnosis of AD. Studies conducted over time indicate that AD patients have decreased activity in the limbic region, particularly the hippocampus, due to brain damage, plaque abnormalities, and cerebral cortex damage. However, these exceptions are less pronounced in MCI patients, indicating that fMRI is less useful for the early detection of MCI. One of the greatest advantages of fMRI is that it does not require the use of radioactive substances; as a result, fMRI can be utilized as often as necessary. Patients with advanced disease stages and severe cognitive impairment are unable to maintain adequate motor control. Therefore, patients must remain still during the scanning process in order to achieve optimal results. The rapid progression of Alzheimer's disease is caused by the neurodegenerative process of intentional connectivity between different brain regions. Various researchers working on resting state fMRI (rs-fMRI) have discussed the presence of common gastrointestinal changes in relation to resting-state systems.

Diffusion Tensor Imaging (DTI)

DTI is an imaging technique based on MRI that depicts minute cross-sectional structural details of brain regions. MRI scanners are used to collect these samples non-invasively. DTI is based roughly on the Browning motion sensation of water molecule activity in human tissues. Consequently, this phenomenon can be described as the microscopic dimension that measures the size, dimensionality, and orientation of the tissue in order to determine the final stage of microscopic degeneration. Studies indicate that DTI can be an implicit method for identifying AD in its earliest stages. The research utilizing DTI-based characteristics may be subdivided into three categories, depending on how the characteristics are extracted: i) tractography, ii) integrated network measurement process, and iii) unique voxel-preference approach.

Positron Emission Tomography (PET)

PET scan is a volumetric subatomic illustration technique used to obtain an anatomical and subanatomical 3D brain scan. For PET scanning, a radioactive isotope is administered or inhaled as a tracing agent, also known as a radiotracer. This serves as an emitter of positrons for the subject. The radiotracer is subsequently detected by a scanning machine. The scanner then produces a digital image (illustration) of the radiotracer's distribution within the subject's body. The nature of the PET scan is dependent on the type of radiotracer utilized. PET scanning has become more expensive due to the use of cyclotron agents, which are essential for the production of radiotracer. As brain function is dependent on blood sugar consumption, it can be deduced from this example that glucose consumption and neural function are directly proportional. Even with mild symptoms, the PET scan is extremely peculiar for predicting AD. The operation of PET scanning is quite efficient, but the aforementioned factors indicate that it is not a healthy diagnostic technique.

Figure 3 shows different brain imaging techniques used in the papers considered. We can see that MRI is the most preferred imaging source with only one paper [19] using DTI along with rs-fMRI

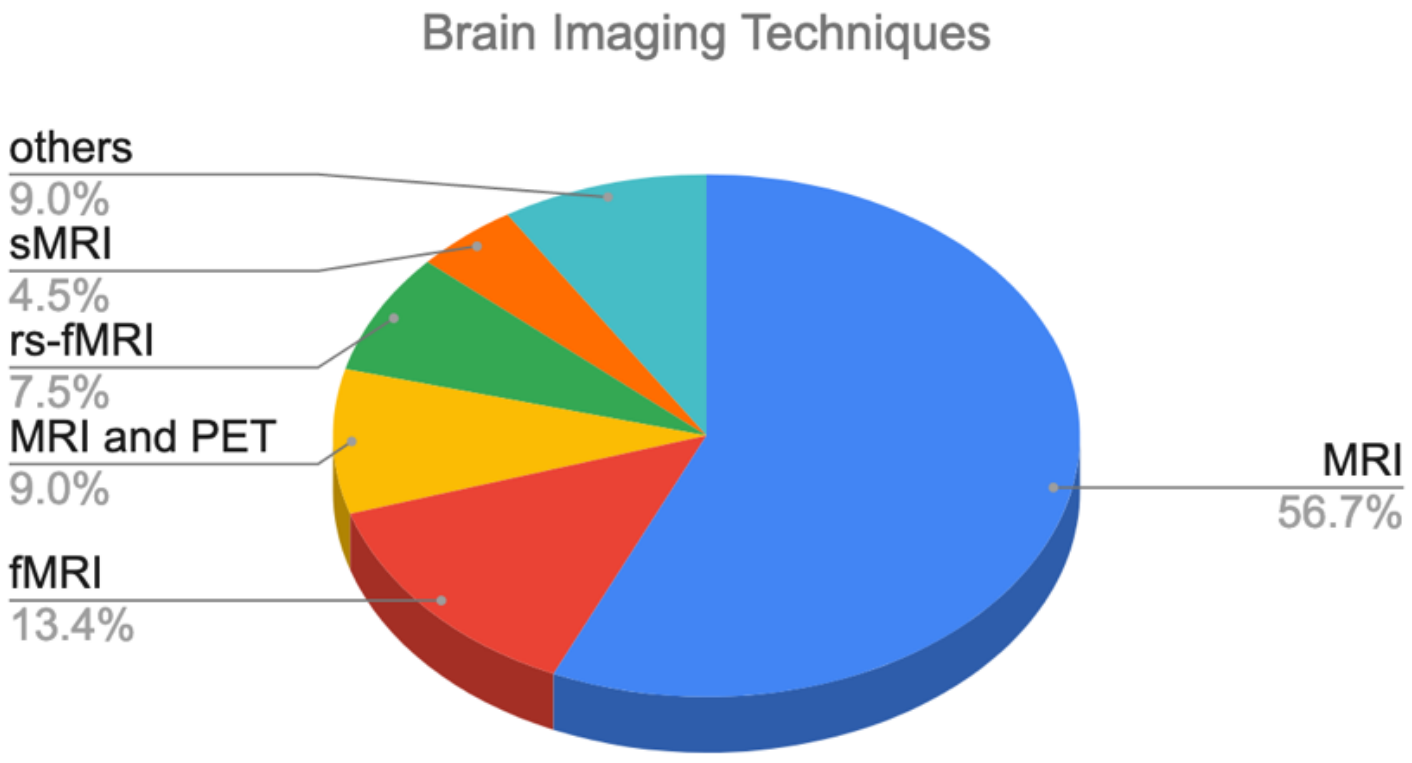

Figure 3: Brain imaging techniques used across reviewed papers.

*Datasets*

Diverse freely online available datasets have fueled the research in automated AD detection. Below, we describe the datasets used by different research papers.

Alzheimer's disease Neuroimaging Initiative (ADNI)

The ADNI [20] is a database accessible at (adni.loni.ucla.edu), that has been used most frequently in studies pertaining to Alzheimer's detection. ADNI was launched in 2003 by the National Institute on Aging (NIA), the Food and Drug Administration (FDA), the National Institute of Biomedical Imaging and Bioengineering (NIBIB), private non-profit organizations, and private pharmaceutical companies.

Open Access Series of Imaging Studies (OASIS)

OASIS is an open-source repository for Alzheimer's disease classification for the scientific community [21]. The compilation of this dataset, which is based on multiple models, is primarily motivated by a desire to facilitate future discoveries in neurodegenerative disease. Formally released OASIS-Cross-sectional and OASIS-Longitudinal data have been applied to hypothesis-driven data analyses, the creation of neuroanatomical atlases, and the development of segmentation algorithms. OASIS is a neuroimaging-based data of longitudinal dimension, clinical dimension data, cognitive-based data, and biomarker dataset. This dataset represents the progression from normal to mild, moderate, and severe stages of the disease, from normal to mild to moderate to severe. These stages can be distinguished by their clinical dementia rating (CDR), which represents the classes as a whole. The OASIS dataset is hosted on a server that gives the community open access to its extensive database of processed MRI images and neuroimaging. This dataset contains extensive demographic, genetic, and cognitive information. This can be used for clinical and cognitive neuroimaging research purposes. The stages of this dataset range from normal ageing to cognitive decline. The OASIS dataset is the foundation for a vast array of research in this field. The dataset is accessible via the website https://www.oasis-brains.org/. The cross-sectional data set included MRI images from 416 subjects ranging in age from 18 to 96 years (young, middle-aged, non-dementia, and older adults with dementia). Three to four T1-weighted scans with a high contrast-to-noise ratio were conducted for each MRI. Here, estimates of total brain volume and intracranial volume were used to examine normal ageing and Alzheimer's disease. The dataset also included information on 20 patients with dementia.

A part of OASIS data is also available at Kaggle (Kaggle Alzheimer’s Classification Data (kACD)) a leading machine learning site for beginners, which hosts a large number of competitions and hackathons related to deep learning. The kACD dataset contains only longitudinal MRI data obtained from 150 subjects aged 60-96.

Besides ADNI and OASIS, one of the considered paper [22] used Multi-Atlas Labelling Challenge (MICCAI 2012) Data [23], and one paper [24] used Minimal Interval Resonance Imaging in Alzheimer's Disease (MIRIAD) [25]. Figure 4 shows the datasets used by the research papers considered in the present review. As reported by earlier review papers, ADNI is still the most widely used dataset, however, more and more researchers are now using OASIS, while in 2021 review the Oasis dataset was only between 2-10%, in the last three years the percentage increased to 26%.

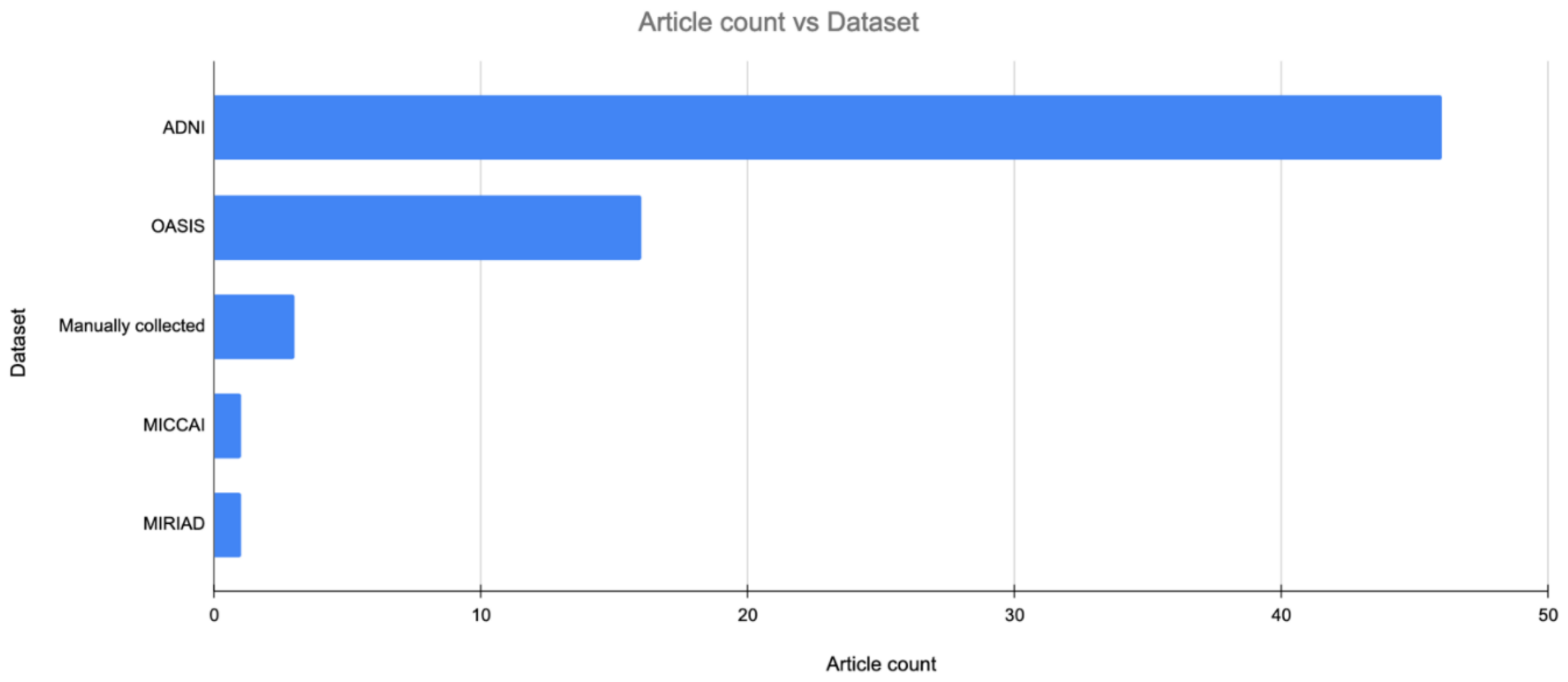

Figure 4: Datasets used by different papers.

*Prediction Models*

Deep learning is one of the most popular techniques in the field of AI today. Deep learning is not one algorithm, but a group of algorithms inspired by the workings of biological neural networks. In this section, we briefly describe some of the deep learning architectures considered for AD detection. The models are categorized as convolutional neural networks based, using transfer learning, hybrid models, and other deep learning and machine learning techniques.

Convolutional Neural Networks and their variants

Convolutional neural networks are deep, feed-forward artificial neural networks well suited for analyzing visual imagery or image data. CNN is composed of multiple layers: one or more convolutional layers followed by a non-linear activation function (typically the ReLU function), and ending with an optional pooling layer, typically max pooling. Additional techniques have been developed to further boost the performance of CNN, including residual learning and dropout. Table 1 lists the research papers and the prediction model used by them. The table also lists the baseline algorithm against which the CNN results are compared.

Table 1 List of Convolutional neural networks and its variant for classification task

| Article | Prediction Model | Baselines | Accuracy# (%) |
|---|---|---|---|
| Liu et al. [26] | S-GCN | Dadi2019, Pdist | 77.2 |
| Salehi et al. [27] | CNN | DSA- 3D CNN, Stacked Autoencoders | 99 |
| Zubair et al. [28] | CNN | N/A | 99.3 (100, 100) |
| Ahmad et al. [28] | CNN | N/A | 97 (96.4,96.4) |
| De Luna et al. [29] | CNN | N/A | 78.9 (79.3,-, 79.3,78.5,78.8) |
| Srivardhini et al. [30] | CNN | N/A | 94 |
| Subaramya et al. [31] | CNN | N/A | 95 |
| Chen et al. [32] | CNN | FCNN, SVM | MMSE (P<0.05), CDR |
| Sethi et al. [33] | 2D-CNN | N/A | 92.55 |
| Goenka et al. [24] | 3D-CNN | N/A | 100 (-,-,100,100,100) |
| Qiao et al. [34] | Ranking CNN (rankCNN) | SVM, 3D CNN, rankCNN-C, rankCNN-CS | RMSE 4.98, CC 0.46 |
| Ramzan et al. [22] | 3D CNN | FreeSurfer+FSL | DSC 0.9 |
| Turkson et al. [35] | Spiking deep CNN | sparse autoencoders and 3D convolutional neural networks, Support Vector Machine (SVM), Random Forest (RF), KNN , Naïve Bayes | 90.2 |
| Amini et al. [37] | Novel convolutional neural network (CNN) | KNN, SVM, DT, LDA, RF | 96.7 (-,-,100) |
| Alshammari et al. [38] | Modified CNN | N/A | 97 |
| Liu et al 2022 [39] | Novel multi-scale CNN (MSCNet) | ResNet-50, ResNet-50+MSRB, MSCNet, MSCNet-U | 90.86 |
| Twaambo et al. [40] | 3D ShuffleNet | ResNet, DenseNet | 85.2 (96, 69, 93.3, 84) |
| Miled et al. [41] | AlexNet | VGG16, GoogleNet, ResNet18 | 96.8 |
| Al-Khuzaie et al. [42] | Alzheimer Network (AlzNet) | N/A | 99.3 (-,-, 98.9, 99.5, 99.2) |
| Sadat et al. [43] | Custom CNN | VGG19, Inception- ResNetv2, ResNet152v2, EfficientNetB5 and EfficientNetB6 | 96 (-,-, 95 ,95, 95) |
| Li et al. [44] | Med-3D network | SVM, RNN, CNN, VGG-16, VGG-19, ResNet | 83 (-,-, 87, 91) |
| Murugan et al. [45] | DEMentia NETwork (DEMNET) | VGG16, Ensemble based Clasifier, Siamese Network | 95.2 (-,-, 96, 95, 95.3) |

# The main metrics is accuracy, in bracket are Specificity, Sensitivity, Precision, Recall and F1 score in respective order.

CNN is conventionally used for image classification problem. From Table 1 we can see that most papers using CNN architecture have defined the problem as classification problem. However, two papers consider it as image segmentation problem, where instead of the entire image, each pixel is classified. For segmentation problems, the Mini-Mental State Examination (MMSE), Dice similarity score and correlation coefficient (CC) are the metrics commonly employed.

Transfer Learning

Transfer learning is used in training deep neural networks with a small amount of training data. It is a method in which a model trained on one task is repurposed for a second task that is related to the first. It results in reduced training time by tuning the pre-trained model on a larger training dataset, using a method known as "fine-tuning". Table 2 lists the articles using the pre-trained models and their respective baselines.

Table 2 List of pre-trained models for classification task

| Article | Prediction Model | Baselines | Accuracy[#] (%) |
|---|---|---|---|
| Naz et al. [46] | VGG | AlexNet, GoogLeNet, VGG-16/19, ResNet-18/50/101, MobileNetV2, InceptionV3, Inception-ResNet-V2 and DenseNet201 | 99.3 |
| Janghel et al. [47] | VGG | SVM, K nearest or linear discriminant classifiers | 100 |
| Mehmood et al. [48] | VGG | N/A | 93.8 (95.1, 92.2) |
| Duc et al. [49] | VGG Net | N/A | 86.4 (100, 67.7) |
| Raju et al. [50] | VGG16 | Inception V4 | 99 |
| Kumar et al. [51] | VGG-16 | N/A | 93.1 (-,-, 100, 100, 93) |
| Rajeswari et al. [52] | VGG-19 | VGG-19, VGG-16, Resnet-50 and Xception | 97 (-,-, 96) |
| Fujibayashi et al. [53] | ResNet | N/A | 87 |
| Puente-Castro et al. [54] | ResNet+SVM | N/A | 86 (97.7,-,73.9, 30.6, 76) |
| Odusami et al. 2021 [55] | Modified ResNet18 | CNN, SVM, BoVM+Late Fusion Scheme | 99.8 (100, 99.6) |
| Buvaneswari et al. [56] | SegNet, ResNet-101 | VGG-16, VGG-19 | 96.3 (96.1, 96.7,98.1) |
| Ebrahimi et al. [57] | ResNet-18 | LeNet-5, AlexNet, VGG-16, SqueezeNet, GooLeNet | 91.8 (92, 91.6) |
| Odusami et al. 2022 [58] | ResNet18 and DenseNet201 | Resnet18 and SVM, 3D CNN, Resnet 18 (Finetuning) | 98.9 (-,-, 98.9, 98.9) |
| Heising et al. [59] | LeNet-5 | 3D CNN | 99 (-,-, 94.8, 89.4, 80.2) |
| Liu et al. [36] | Depthwise Separable Convolution | AlexNet, GoogleNet | 77.79 |
| Ashraf et al. [60] | DenseNet | AlexNet, GoogLeNet, VGG-16, Inception-ResNet-v2, ResNet-18/50/101, MobileNetV2, InceptionV3, Inception-ResNet-V2 and DenseNet201 | 99.05 |
| Fiasam et al. [61] | Enhanced ResNet-18 | 3D CNN, 3D ResNet | 92.12 (88.2, 86.1) |

# The main metrics is accuracy, in bracket are Specificity, Sensitivity, Precision, Recall and F1 score in respective order.

Transfer learning has made it possible to use Deep Networks even when data is not sufficient. In almost all image classification tasks, leveraging transfer learning improves performance. Our review also shows that compared to scratch CNNs, the model using transfer learning approach are performing better, with best results from DenseNet.

Hybrid Models

Beside CNN, some other models employed for AD detection include methods like Principal component Analysis, support vector machine, autoencoder, k-means and genetic algorithms. Table 3 lists research papers with hybrid models, that is they use more than one type of algorithms, mostly a deep learning algorithm along with a machine learning algorithm.

Table 3 List of hybrid models between deep learning models and traditional models for classification task

| Article | Prediction Model | Baselines | Accuracy# (%) |
|---|---|---|---|
| Bi et al. [62] | PCANet + k -means | 3D CNN, Stacked auto encoders | 97.01 |
| Jia et al. [63] | 3DPCANet | N/A | 95 (100, 100, -,- ,95.6) |
| Wang et al. [64] | 3DPCANet+SVM+mALFF | 3DPCANet model | 91.8 (92.5, 94.3, _,_, 87.4) |
| Mohammed et al. [65] | AlexNet+SVM and ResNet-50+SVM | SVM, Decision Tree, Random Forest, KNN | 94.8 (97.8, 93) |
| Ahmadi et al. [66] | Sparse Autoencoder + Genetic Algorithm + SVM | N/A | 98.35 |

# The main metrics is accuracy, in bracket are Specificity, Sensitivity, Precision, Recall and F1 score in respective order.

Other Models Used for prediction

Table 4 lists other deep learning-based algorithms and machine learning methods employed for AD detection. These include models based on Attention mechanism. Attention based models have become very popular in natural language-based tasks. Motivated by their success in NLP, researchers have also used it in image-based tasks. Another notable algorithm is Capsule Network, first proposed by Sabour, Fross, and Hinton [70], increases the weights of similar information through its dynamic routing and replaces the pooling operation employed by conventional convolution neural networks.

Table 4 List of other types of models for classification task

| Article | Prediction Model | Baselines | Accuracy# (%) |
|---|---|---|---|
| Bi et al. [68] | Extreme Learning Machine | N/A | 61.3 |
| Wang et al [69] | Hierarchal Extreme Machine learning | N/A | 88.7 (79, 83) |
| Devika et al. [70] | ANOVAF + SVM | N/A | 80.8 |
| Sadiq et al. [69] | minimum-redundancy maximum-relevance (mRMR) + ReliefF algorithm + SVM | 3DCNN, Elastic Net | 96.36 |
| Hsieh et al. [70] | Behavior score-embedded encoder network (BSEN) | ICA, PCA, CAE(3D convolutional autoencoder without contrastive loss embedding) | UAR 59.4 |
| Liu et al. 2020a [72] | Brain functional connectivity network (BFCN) + attention-based bidirectional LSTM (ASBiLSTM) | PC+ASBiLSTM, SSN+A SBiSR, SSN+SBiLSTM | 91.5 (88.9, 96.6) |
| Saratxaga et al. [73] | BrainNet2D | Inception, Xception, ResNet | BAC 93 |
| Ali et al. [74] | Deep Convolutional Second-Generation Curvelet Transform Network (SGCTN) | Original Features Curvelet Network (OFCN) | 97.3 (96.1, 98.8) |
| Zhang et al. [19] | Deep cross-model attention network (DCMAT) | GCRNN, Sparse Autoedcoder with CNN, CNN | 98.6 |
| Wang et al. [64] | Deep fusion model: integrates autoencoder, multi-class classification and structure learning | N/A | 73 (NC/LMCI/EMCI) |
| Hedayati et al. [75] | Ensemble Convolutional Auto Encoder-Decoder (ECAED) | 3D CNN, Stacked Bidirectional RNN, Ensemble CNN | 95 (100, 92) |
| Fang et al. [76] | Ensemble DCNNs of multi-modality images by Adaboost learner (DCMA) | GoogLeNet, ResNet, DenseNet, Ensemble, D-Ensemble | 99.27 (98.72, 95.89) |
| Liu et al. 2020b [77] | GCN-EMCI | N/A | 84.1 (81.3, 86.5) |
| Haouas et al. [78] | Mamdani (Fuzzy Logic) | 3D Autoencoder, Multi-Modal Deep Neural Network | 99.1 |
| Baskaran et al. [79] | GAN | | 96 |
| Basheer et al. [80] | MCapNet model (Capsule Network) | Naïve Bayes Models, SVM, Decision Trees, XGBoost, Gradient Boost, Ensemble, Adaboost, Random forest models | 92.3 (-,-,-,82.3,88.8) |
| Fan et al. [81] | U Net | VoxNet+ResNet, K-NN, DNN, CNN, ELM, DBN, SRNN, D2ML | 95.71 |

# The main metrics is accuracy, in bracket are Specificity, Sensitivity, Precision, Recall and F1 score in respective order.

The results from the above approaches are interesting, more so because they attempt combine feature selection procedures with classification. Additionally, many of the above papers are attempting 3 class or more, with an emphasis on improving early MCI (EMCI) detection.

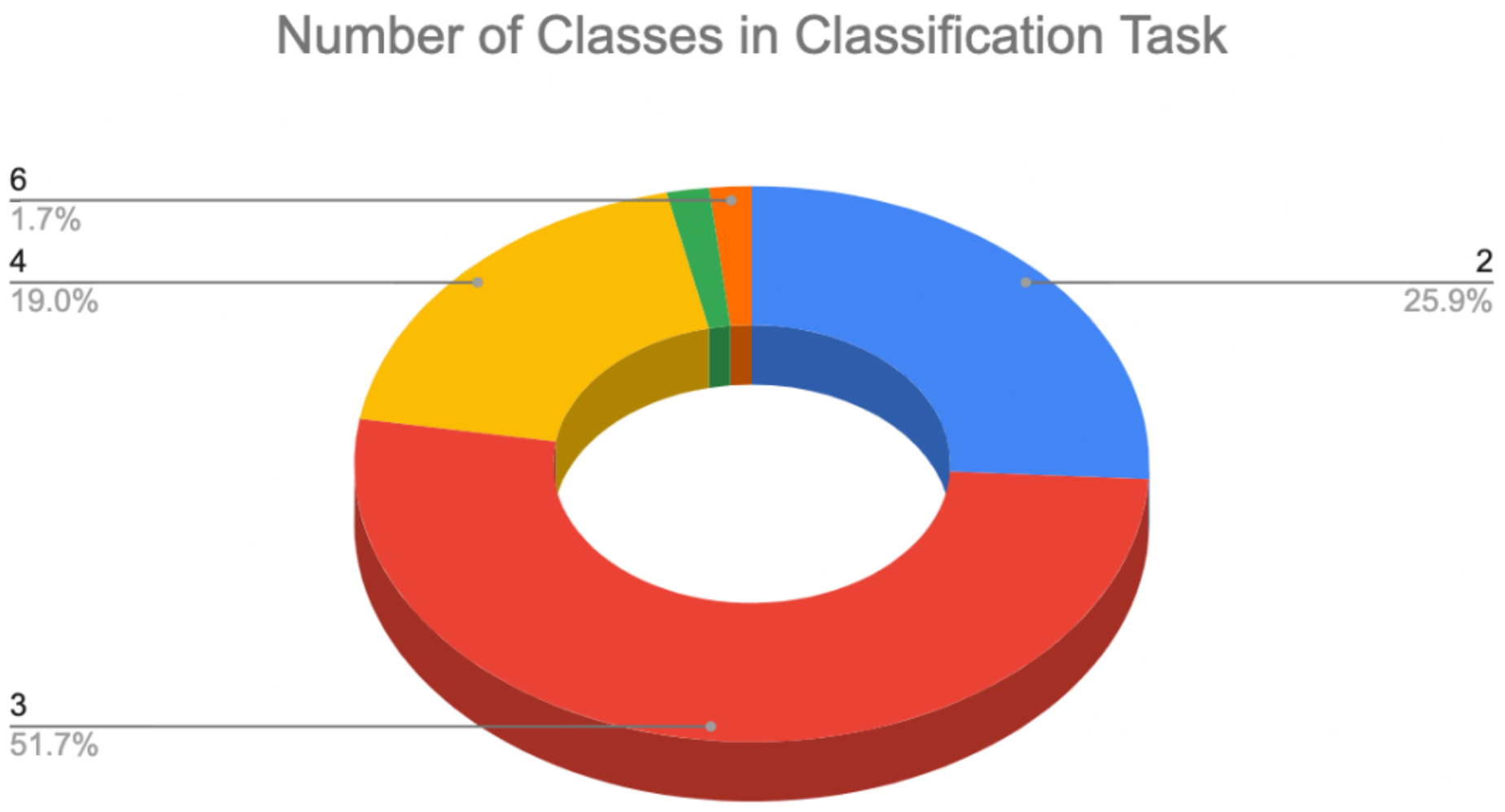

Figure 5: The distribution of articles based on the number of classes in the classification task

Figure 5 shows the distribution of the articles according to the number of classes in the classification task. Most research papers explored three classes, though there are few trying to attempt up to six classes.

*Model evaluation*

There are several metrics for measuring the performance of various machine learning algorithms. The metrics like accuracy, sensitivity, specificity, precision, F1 Score, ROC, and AUC are used in the classification tasks.

1. Accuracy: Accuracy is defined as the ratio of correct predictions to the total number of predictions. The mathematical expression for accuracy is given as:

$$\text{Accuracy} = \frac{t^P + t^N}{t^P + t^N + f^P + f^N}$$

2. Sensitivity/ True Positive Rate (TPR): Sensitivity is defined as the proportion of positive responses correctly identified by the classifier. The expression of TPR is given by

$$\text{Sensitivity} = \frac{t^P}{t^P + f^N}$$

3. Specificity/ True Negative Rate (TNR): Specificity is defined as the proportion of the negative responses correctly identified by the classifier. Mathematically, it can be written as

$$\text{Specificity} = \frac{t^N}{t^N + f^P}$$

4. Precision: Precision is defined as the percentage of true positives among all positive results. Mathematically, we can express it as:

$$\text{Precision} = \frac{t^P}{t^N + f^P}$$

5. F1 Score: It defines the average response of the precision and the sensitivity metrics. The expression of F1 score in terms of sensitivity and precision is

$$F1\ score = \frac{\text{Sensitivity} \ * \ \text{Precision}}{\text{Sensitivity} + \text{Precision}}$$

6. Receiver Operating Characteristics (ROC) curve: At a certain threshold, it's easy to figure out how well a classifier can separate two groups. The ROC curve, which combines sensitivity and specificity from the method, is easy to understand and is not affected by changes in how the classes are distributed. It accurately shows the relationship between sensitivity and specificity of a certain analytical method and gives a full picture of how accurate a diagnostic method is.
7. AUC: The area under the ROC curve (AUC) is a simple way to judge how good a classifier is; the higher the value, the better. The model is more accurate if the ROC curve is close to the upper left corner. The best classification threshold is the point on the ROC curve that is closest to the upper left corner. The more accurate the classification algorithm is, the more likely it is that the positive samples will be put ahead of the negative samples if the AUC value is high.
8. R-Squared (R2): R squared, also called the coefficient of determination, is a statistical measure that tells you how well a line fits the data. It is used in regression analysis to determine how much of the variability in the dependent variable can be explained by the independent variable. The higher the R squared value, the better the fit. A value of 0 means that the line does not fit the data at all, while a value of 1 means that the line fits the data perfectly. R squared can be negative if the line actually fits the data better in the reverse direction. The closer R squared is to 1, the better the line fit. The R square value can be calculated using the following formula:

$$R^2 = 1 - \frac{\text{SSE (Sum of squared errors)}}{\text{SST (Sum of squared deviations from the mean)}}$$

9. Root Mean Square Error (RMSE): In machine learning, the root mean square error (RMSE) is a measure of the difference between predicted values and actual values. The RMSE measures the average deviation of all predictions from the actual value. It is a measure of how accurately a model predicts the response variable.
10. Mean Absolute Error (MAE): The mean absolute error is a measure of how close predictions are to actual values. It is calculated as the average of the absolute difference between predictions and actual values.
11. Mini-Mental State Examination (MMSE): It estimates the severity of cognitive impairment in seven categories: orientation to time, orientation to place, registration of three words, attention and calculation, recall of three words, language, and visual construction. The total score is 30 points.
12. Dice Similarity Coefficient: The Dice similarity coefficient is a statistical measure of how similar two sets of data are. The coefficient is calculated by taking the number of shared items divided by the total number of items.
13. Positive Predictive Value: Positive predictive value (PPV) is the probability that a person with a positive test result actually has the disease. In other words, it is the likelihood that a positive test result is accurate. PPV is affected by the sensitivity and specificity of the test, as well as the prevalence of the disease in the population. A high PPV means that there is a greater chance that a positive test result is accurate, while a low PPV means that there is a greater chance that a positive test result is not accurate.
14. Correlation coefficient: The correlation coefficient is a statistical measure that calculates the strength of the relationship between two variables. This value can range from -1 to 1, with -1 indicating a strong negative correlation and 1 indicating a strong positive correlation.

The detailed list of research papers using different metrics is shown in Table 11.

Table 5 Article Lists archived by different evaluation metrics.

| Metrics | Article List | Range |
|---|---|---|
| Accuracy | Duc et al. [49], Bi et al. [62, 67], Salehi et al. [27], Zhang et al. [19], Puente-Castro et al. [54], Liu et al. [26, 36, 39, 72, 77], Fang et al.[76], Ahmadi et al. [66], Miled et al. [41], Wang et al. [64, 68], Murugan et al. [45], Al-Khuzaie et al.[42], Naz et al. [46], Janghel et al. [47], Ashraf et al. | 57.26% - 100 % |
| | [60], Goenka et al. [24], Devika et al. [70], Amini et al. [37], Zubair et al. [27], Ahmad et al. [28], Turkson et al. [35], De Luna et al. [29], Sethi et al. [33], Alshammari et al. [38], Buvaneswari et al. [56], Sadat et al. [43], Subaramya et al. [31], Mohammed et al. [65], Li et al. [44], Twaambo et al. [40], Fujibayashi et al. [53], Baskaran et al. [79], Sadiq et al. [69], Mehmood et al. [48], Saratxaga et al. [73], Ali et al. [74], Fan et al. [81], Kumar et al. [51], Hedayati et al. [75], Odusami et al. [55, 58], Fiasam et al. [61], Heising et al. [59], Basheer et al. [80], Jia et al. [63] | |
| Precision | Puente-Castro et al. [54], Murugan et al.[45], Al-Khuzaie et al. [42], Goenka et al. [24], Amini et al. [37], De Luna et al. [29], Jia et al. [63], Buvaneswari et al. [56], , Sadat et al. [43], Rajeswari et al. (2021), Mohammed et al. [65], Li et al. [44], Basheer et al. [80], Twaambo et al. [40], Chen et al. [32], Kumar et al. [51], Wang et al. [64], Odusami et al. [58], Heising et al. [59] | 70.64% - 100% |
| Recall | Puente-Castro et al. [54], Bi et al. [67], Murugan et al. [45], Al-Khuzaie et al.[42], Goenka et al. [24], De Luna et al. [29], Sadat et al. [43], Rajeswari et al. (2021), Mohammed et al. [65], Liu et al. [36], Basheer et al. [80], Twaambo et al. [40], Kumar et al. [51], Odusami et al. [58], Heising et al. [59], Jia et al. [63] | 28.22% - 100% |
| Specificity | Duc et al. [49], Bi et al. [62], Puente-Castro et al. [54], Liu et al. [36, 39, 72, 77], Fang et al. [76], Janghel et al. [47], Ashraf et al. [60], Zubair et al. [27], Odusami et al. [58], Ahmad et al. [28], Wang et al. [64, 68], De Luna et al. [29], Buvaneswari et al. [56], Mohammed et al. [65], Twaambo et al. [40], Mehmood et al. [48], Ali et al. [74], Hedayati et al. [75], Fiasam et al. [61], | 79.3% - 100% |
| Sensitivity | Duc et al. [49], Bi et al. [62], Liu et al. [36, 39, 72, 77], Fang et al. [76], Janghel et al. [47], Ashraf et al. [60], Amini et al. [37], Zubair et al. [27], Odusami et al. [58], Ahmad et al. [28], Wang et al. [64, 68], Buvaneswari et al. [56], Mohammed et al. [65], Twaambo et al. [40], Mehmood et al. [48], Ali et al. [74], Hedayati et al. [75], Fiasam et al. [61] | 69% - 100% |
| F1 score | Puente-Castro et al. [54], Al-Khuzaie et al.[42], Goenka et al. [24], De Luna et al. [29], Jia et al. [63], Sadat et al. [43], Mohammed et al. [65], Basheer et al. [80], Twaambo et al. [40], Kumar et al. [51], Wang et al. [64], Heising et al. [59] | 39.91% - 100% |
| ROC | Liu et al.[77], Goenka et al. [24], Amini et al. [37], Jia et al. [63], Basheer et al. [80] | - |
| AUC | Liu et al. [26, 36, 72, 77], Bi et al.[67], Fang et al. [76], Murugan et al.[45], Goenka et al. [24], Amini et al. [37], Wang et al. [64, 68], Jia et al. (2021), Mohammed et al. [65], Twaambo et al. [40], Fujibayashi et al. [53], Fan et al. [81] | 69.9% - 100% |
| R2 (R-squared) | Duc et al. [49]] | 0.63 ± 0.02 |
| Root mean square error (RMSE) | Duc et al. [49] | 3.27 ± 0.58 |
| MMSE | Duc et al. [49], Bi et al. [62], Hsieh et al. [70], Chen et al. [32] | 55.11 |
| Dice similarity coefficient (DSC) | Chen et al. [32] | 0.9 |
| MAE | Ali et al. [74], Chen et al. [32] | 5.582 ± 0. 012 |
| Correlation coefficient (CC) | Qiao et al. [34] | 0.288 ± 0.003 |

## 5. Implementation and Reproducibility

In this section, we pay a special attention to the implementation details of the papers we survey, which is less discussed before in previous surveys. The programming language used to implement machine learning and deep learning models is investigated first. Python has become the dominant choice for model implementation over the past three years. It offers a variety of packages and frameworks, such as TensorFlow [82, 83, 84], PyTorch [85], and scikit-learn [86, 87, 88]. Other options include R, MATLAB, and Java.

*Computational Resources*

Deep learning models require a larger amount of computation for training, and GPUs have been used to accelerate the convolutional operations involved. With the need to process multiple types of input data, especially text data, the need for GPU will keep increasing in this research area. In Table 6, we give a list of different types of GPUs used in the surveyed papers.

Table 6 List of GPUs used for Alzheimer disease classification

| GPU Type | Article List |
|---|---|
| GPU NVIDIA GeForce GTX 1070 Ti | Kumar et al. [51] |
| GTX 1070 GPU | Ashraf et al. [60] |
| Intel Core i3 2.7 GHz CPU, 6GB RAM | Wang et al. [64] |
| Intel Core i7-8550 U, RAM 16 GB DDR4, 250 GB SSD hard. | Ahmadi et al. 2020 [66] |
| Inter core i7, 8 GB memory and GPU with 16G NVIDIA P100 × 8 | Fang et al. [76] |
| Intel i5 8th Gen processor | Basheer et al. [80] |
| NVIDA Tesla K80 GPUs | Heising et al. [59] |
| NVIDIA Corporation Tu116 (Geforce GTX 1660) GPU | Odusami et al. [58] |
| NVIDIA DXG-1 | Liu et al. 2021 [36] |
| NVIDIA DXG-1, Tesla P100 GPU | Liu et al. 2022 [39] |
| NVIDIA GeForce GTX 1080 Ti | Duc et al. [49], Bi et al. [62], Qiao et al. [34] |
| NVIDIA GeForce RTX 3060 GPU | Fiasam et al. [61] |
| NVIDIA Quadra RTX6000 workstation with a 24GB GPU | Murugan et al. [45] |
| NVIDIA RTX2070 GPU | Jia et al. [63] |
| NVIDIA Tesla K80 GPU | Subaramya et al. [31] |
| NVIDIA TITAN Xp GPU | Fan et al. [81], Twaambo et al. [40], Ali et al. [74] |
| NVIDIA V100 GPU | Ebrahimi et al. [57] |
| NVIDIA Volta GPU | Goenka et al. [24] |
| ZOTAC 11GB GPU | Ramzan et al. [22] |

*Code Availability*

GitHub has been the mainstream platform for hosting source code in the computer science field. In

Table 7, we list the articles with public code repositories.

Table 7 List of articles with public source code links.

| Article | Public Source Code Link |
|---|---|
| Puente-Castro et al. [54] | https://github.com/TheMVS/DL_AD_mri_sex_age_stages |
| Ashraf et al. [60] | https://github.com/imranrazzak/Neurological-Disorder |
| Turkson et al. [35] | https://github.com/mvisionai/AdSpike |
| Fan et al. [81] | https://github.com/ETVP/AD-diagnosis |
| Qiao et al. [34] | https://github.com/fengduqianhe/ADrankCNN-master |

## 6. Conclusion and Discussion

Neuroimaging studies have demonstrated that classification algorithms and neuroimaging can be used to detect AD in the prodromal phase prior to the onset of clinical symptoms. Numerous research institutions are therefore committed to the classification of AD and MCI based on neuroimaging, employing various neuroimaging techniques such as sMRI, fMRI, and PET. The characteristic changes in the brains of AD and MCI patients can aid in controlling the pathophysiology of AD. Utilizing deep learning to classify AD has made significant progress. Recent trend in research of using end to end deep learning methods present great potential in the early detection of MCI. Since structural changes appear prior to the onset of the clinical symptoms, there is a valuable window of time during which the morphological and functional changes in the brain can be detected and, hence, used to predict and provide clinical treatment to slow down the progression of Alzheimer.

It was challenging to compare different review papers because of highly variable methodologies including different datasets, preprocessing methods, inputs selected and also the final task of classification or segmentation.

Regarding the review's limitations, it is important to note that we omitted methodological details such as the preprocessing methods used to obtain neuro-imaging results and the mathematical development of the algorithms. This information could have provided a better understanding of the performance of each model and how the data is classified to differentiate between groups, but these in-depth methodological analyses were beyond the scope of this review due to its focus on deep learning techniques.

The future result appears to be moving in the direction where more information can be fed to deep neural networks using techniques like spatio-temporal maps, structural brain network, or multimodal models where beside fMRI images, text and voice data is also included. To enable clinical integration of the models, the work on approaches to reduce model size while not effecting the performance can be useful.

New and increasingly complicated deep learning architectures will continue to be created, and access to higher levels of processing capacity as well as the precision and resolution of neuroimaging techniques will continue to increase. Therefore, in the future, faster, more precise, and more efficient classification systems may be directly included into neuroimaging techniques, enabling the production of a diagnostic hypothesis from a single brain scan.

**Author Contributions:** Conceptualization, Amita Kapoor. and Narotam Singh; methodology, Narotam Singh.; software, Narotam Singh.; validation, Pateshwari D., Narotam Singh and Amita Kapoor; formal analysis, Neha Soni; data curation, Neha Soni; writing—original draft preparation, Amita Kapoor; writing—review and editing, Narotam Singh and Amita Kapoor.; All authors have read and agreed to the published version of the manuscript

**Funding:** This research received no external funding

**Data Availability Statement:** Not applicable.

**Conflicts of Interest:** The authors declare no conflict of interest".